\documentclass[aps,preprintnumbers,amsmath,twocolumn,groupedaddress,superscriptaddress,notitlepage,nofootinbib,prd]{revtex4-1}
\usepackage[utf8]{inputenc}
\usepackage{indentfirst}
\usepackage{amssymb}
\usepackage{amsmath,latexsym}
\usepackage{graphicx}
\usepackage{float}
\usepackage[margin=1.0in]{geometry}
\usepackage[mathscr]{euscript}
\usepackage{color}
\usepackage[dvipsnames]{xcolor}
\usepackage{soul}
\usepackage{slashed}
\usepackage{feynmf}
\usepackage{subcaption}
\usepackage{braket}
\usepackage{ulem}
\usepackage{comment}
\usepackage{dsfont}

\newcommand{\beq}{\begin{equation}}
\newcommand{\eeq}{\end{equation}}
\newcommand{\bea}{\begin{eqnarray}}
\newcommand{\eea}{\end{eqnarray}}

\long\def\beqs#1\eeqs{\beq\begin{split} #1 \end{split}\eeq}

\definecolor{MyRed}{RGB}{153,0,13}

\usepackage[colorlinks=true,backref=true,linktocpage=true,citecolor=MyRed,urlcolor=MyRed,linkcolor=MyRed,pdfpagemode=UseOutlines]{hyperref}

\hypersetup{%
  bookmarksnumbered=true,
  pdftitle = {},
  pdfsubject = {},
  pdfauthor = {},
  pdfkeywords = {}
}

\usepackage{microtype}

\usepackage{pgfplotstable,array,siunitx}
\usepackage{multirow}
\usepackage{booktabs}

\usepackage{tikz}

\preprint{INT-PUB-20-030}

\begin{document}
\title{Neural Network Solutions of Bosonic Quantum Systems in One Dimension}

\author{Paulo F. Bedaque}
\email{bedaque@umd.edu}
\affiliation{Department of Physics,
University of Maryland, College Park, MD 20742}

\author{Hersh Kumar}
\email{hekumar@umd.terpmail.edu}
\affiliation{Department of Physics,
University of Maryland, College Park, MD 20742}

\author{Andy Sheng}
\email{asheng@umd.edu}
\affiliation{Department of Physics,
University of Maryland, College Park, MD 20742}

\preprint{}

%\date{Aug. 2023}
\pacs{}

\begin{abstract}

%Understanding a many-body quantum system from first principles typically involves solving a Schrödinger equation of high complexity. In many cases, analytic solutions are not accessible and approximations must be sought. 

Neural networks have been proposed as efficient numerical wavefunction ansatze which can be used to variationally search a wide range of functional forms for ground state solutions. These neural network methods are also advantageous in that more variational parameters and system degrees of freedom can be easily added. We benchmark the methodology by using neural networks to study several different integrable bosonic quantum systems in one dimension and compare our results to the exact solutions. While testing the scalability of the procedure to systems with many particles, we also introduce using symmetric function inputs to the neural network to enforce exchange symmetries of indistinguishable particles.

\end{abstract}

\maketitle

\section{Introduction}

Solving many-body quantum mechanical systems involves obtaining solutions to a high-dimensional partial differential equation or diagonalizing exponentially large Hamiltonians, both of which pose significant challenges.
%Solving a many-body quantum mechanical system poses significant challenges. In particular, for systems of many particles in continuous space, the time-independent Schrödinger equation becomes a high-dimensional partial differential equation, making it difficult to find solutions given general interaction potentials. For spin systems, diagonalizing the Hamiltonian becomes difficult since the Hilbert space is exponentially large in the number of particles. 
Thus, analytically solved models are few and far between. \cite{PhysRev.130.1605, PhysRev.130.1616, PhysRevB.38.6689, 10.1063/1.1664821, PhysRevA.4.2019, MOSER1975197, PhysRevLett.125.220602}.\\

Various methods have been developed to obtain approximate solutions to complex quantum systems. For one such method called Variational Monte Carlo, a variational ansatz for the ground state wavefunction is assumed and an estimate for the ground state energy is obtained by minimizing the energy of the system over the parameters of the chosen approximate wavefunction form. The use of trial wavefunctions has been used to study a broad range of many-body systems, from Fermi and Bose gases to quantum chemistry \cite{Fetter_1997, PhysRevB.18.3126, WILSON20141065, PhysRevLett.60.1719, RevModPhys.73.33, PhysRevB.92.245106, Kościk_2018, PhysRev.128.1791, CARLSON198359, carlson1987microscopic}. In particular, the Slater-Jastrow \cite{PhysRev.98.1479} wavefunction ansatz has been employed widely \cite{lutsyshyn2017weakly, article, Kościk_2020} because, by construction, it captures correlations between interacting particles and incorporates the correct intuitive functional form of the physical system. Tensor networks and density matrix renormalization group methods \cite{PhysRevLett.69.2863} were shown to efficiently encode low-lying eigenstates of local, gapped Hamiltonian spin systems using only a polynomial number of parameters. This eliminates the need to search through an exponentially-large Hilbert space. \\

More recently, established machine learning techniques such as neural networks have found their way into quantum physics \cite{Das_Sarma_2019, RevModPhys.91.045002}. In the context of studying ground states of quantum systems, neural networks conveniently provide a numerical wavefunction ansatz for which one can vary to minimize the system energy cost function. The minimization can be done efficiently using established machine learning methods of backpropagation and gradient descent. Another appeal of using neural networks is that they were shown to be universal function approximators \cite{HORNIK1989359, HORNIK1991251} and thus can be used to describe any general wavefunction. As a consequence, the number of variational parameters is not intrinsically tied to the function form (i.e. Jastrow) of the ansatz and can be increased easily. \\

Carleo and Troyer \cite{Carleo_2017} showed that a type of neural network called restricted Boltzmann machines were sufficient to describe ground state wavefunctions of spin systems with a reasonable number of parameters. Later, restricted Boltzmann machines was found to map directly to matrix product state approximations of spin states \cite{PhysRevB.97.085104, PhysRevB.99.155129}. For particles interacting via continuous-space potentials, several systems of indistinguishable particles have also been studied using neural networks including the Calogero-Sutherland model \cite{Saito_2018}, nuclear models \cite{Gnech_2021, PhysRevLett.127.022502, PhysRevResearch.4.043178}, and chemical systems \cite{nordhagen2022efficient, Pfau_2020}. \\

Quantum systems involving indistinguishable particles must have ground state wavefunctions which satisfy position exchange (anti)symmetries. This poses a challenge in which the constructed neural network trial wavefunction should satisfy these constraints. In one spatial dimension, Bose symmetry can be satisfied by feeding in ordered coordinates of the bosons into the neural network \cite{Saito_2018}; however, this method is not easily generalizable to systems of higher dimensions. Pfau et. al. \cite{Pfau_2020} developed the FermiNet neural network structure to impose fermionic antisymmetry to study electron-electron and electron-ion interactions in atoms. The FermiNet structure constructs the most general fermionic wavefunction space for which one can search for the ground state through by taking Slater determinants of permutation-equivariant functions. The permutation-equivariance of such functions are enforced through permutation-equivariant neural network layers. \\

%Although the FermiNet structure constructs the most general fermionic wavefunction space for which one can search for the ground state through, it enforces the anti-symmetry via layer by layer operations in an additional neural network. This additional structure adds extra complexities to explore, in particular, how the generalization of the wavefunction depends on the number of layers and parameters in that structure. \\

In this paper, we instead propose the use of carefully selected symmetric inputs to enforce the Bose symmetry of the neural network ansatz. By using symmetric functions as inputs to the neural network, we are able to construct the most general Bose symmetric function as we increase the number of nodes and layers; this gives us full flexibility in changing the architecture of the network. We benchmark the use of symmetric functions by computing ground state energies and wavefunctions of several one-dimensional, exactly-solvable quantum systems and comparing both quantitative and qualitative results. Our methodology is not restricted to studying exactly-solvable systems and we also explore systems in which no analytical solution is known. We further demonstrate the scalability and versatility of our neural network ansatz by calculating ground states of systems with dozens of particles. \\

In Sec. \ref{models}, we outline the various models for which we study and compare numerical results to. In Sec. \ref{methods}, we explicitly construct the Bose symmetric neural network functions and outline our procedure for computing ground states. Sec. \ref{results} discusses our results; we find that our neural network ansatz is able to accurately compute ground state energies and features of the many-body systems considered over their various phases. Then we conclude with Sec. \ref{conclusions} by exploring prospects of using symmetric function constructions to write general bosonic and fermionic trial wavefunctions in a higher number of spatial dimensions.

\section{Models \label{models}}

To show the validity of our symmetric input neural network solutions, we apply our neural-network-based methodologies to solve for the ground state wavefunctions and energies for several different one-dimensional quantum systems. We consider both a system of cold bosons in a harmonic trap and a system of trapped bosons with short and long-range interactions. Both quantum systems are inspired by the Lieb-Liniger model \cite{PhysRev.130.1605, PhysRev.130.1616} --- a homogeneous gas of indistinguishable bosons interacting via a contact delta-function potential which was shown to be exactly solvable in one dimension through the use of Bethe ansatz \cite{PhysRevLett.19.1312, PhysRevA.2.154, 1969JMP....10.1115Y, 10.1063/1.1704156}. These Lieb-Liniger-inspired models have rich phase structure and have been studied through the lenses of both finding exact solutions and performing numerical analyses \cite{busch, WILSON20141065, Minguzzi:2022bne, PhysRevLett.125.220602, Beau:2021kso, Yang:2022bfk}. Thus, they provide good testbeds for which to compare both quantitative and qualitative ground state results to.  

\subsection{Cold Bosons in a Harmonic Trap \label{coldbosons}}

The first Hamiltonian which we consider describes spinless bosons in a harmonic trap. The interactions of the condensate are approximated by a contact potential. Using units where $\hbar = 1$, the Hamiltonian is
\begin{equation}
    \hat{H} = \sum_{i=1}^{N} \left(-\frac{1}{2m}\frac{\partial^2}{\partial x_i^2} + \frac{1}{2}m\omega^2x_i^2\right) + \sum_{i < j}^{N} g\delta(x_i - x_j) \label{gpmodel}
\end{equation}
In general, we take $m = \omega = 1$. The interaction is repulsive when $g > 0$ and attractive when $g < 0$. When $g > 0$, the interaction strength $g$ is proportional to the two-boson s-wave scattering length \cite{busch}. Although the addition of the harmonic trap causes the system to no longer be analytically solvable \cite{WILSON20141065} for the general case of many particles, the two-boson case has been solved exactly \cite{busch, WILSON20141065}. 

\subsection{Trapped Bosons with Short and Long-Range Interactions \label{longrangeint}}

Recently, another quantum system has been shown to be exactly solvable for a general case of $N$ bosons \cite{PhysRevLett.125.220602}. The model consists again of particles in harmonic confinement interacting with a delta-function potential of strength $g$. Additionally, a long-range linear interaction potential of strength $\sigma$ is added, which, in one dimension, corresponds to either Coulomb repulsion ($\sigma < 0$) or gravitational attraction ($\sigma > 0$).
\begin{equation}
\begin{split}
    \hat{H} &= \sum_{i=1}^{N}\left(-\frac{1}{2m}\frac{\partial^2}{\partial x_i^2} + \frac{1}{2}m\omega^2x_i^2\right) \\
    &+ \sum_{i < j}^N\left(g\delta(x_i-x_j) + \sigma|x_i-x_j|\right) \label{longham}
\end{split}
\end{equation}
Again, we take $m = \omega = 1$. The model was found to be integrable when $\sigma = -m\omega g/2$\footnote{References \cite{PhysRevLett.125.220602, Yang:2022bfk} claim the solution found is for the regime where $\sigma = -m\omega g$. However, we analytically find that the provided solution is correct not for $\sigma = -m\omega g$, but for $\sigma = -m\omega g/2$.} \cite{Yang:2022bfk}. In that case, $g > 0$ corresponds to both interaction terms being repulsive and $g < 0$ corresponds to both terms being attractive. The coupling strength is again related to the one-dimensional s-wave scattering length $a_s = -2/(mg)$. The exact expression for the ground state energy of the Hamiltonian \eqref{longham} was found to be
\begin{equation}
    E_0 = \frac{N\omega}{2} - mg^2\frac{N(N^2-1)}{24} \label{gse}
\end{equation}
with a corresponding (non-normalized) ground state wave function 
\begin{equation}
    \psi_0(x_1,...x_N) = \prod_{i<j}e^{-|x_i-x_j|/a_s}\prod_i e^{-x_i^2/(2a_{\text{ho}}^2)} \label{gswave}
\end{equation}
where $a_{\text{ho}} = \sqrt{1/(m\omega)}$ denotes the harmonic oscillator length. \\

When $g > 0$, the ground state wavefunction is maximized at $|x_i - x_j| = Na_{\text{ho}}^2/a_s$ and the bosons tend to spread out with equal spacing about the center of the harmonic trap. When a weak repulsion is turned on, the system forms an incompressible liquid of flat density \cite{PhysRevLett.50.1395, lieb2018rigidity}. In the limit of strong repulsion $g \gg 1$ the probability of overlap between any two particles drops to zero and the system forms a Wigner crystal \cite{PhysRev.46.1002}. In the other regime, when $g < 0$, the bosons collapse into the center of the trap. In the limit of $g \rightarrow -\infty$, the ground state wavefunction of \eqref{longham} becomes 
\begin{equation}
    \psi_0(x_1,...,x_N) = \prod_{i<j}e^{-|x_i-x_j|/a_s}
\end{equation}
resembling a McGuire bound-state solution \cite{10.1063/1.1704156} describing a bright soliton \cite{inbook}. It is surprising that despite the fact that different signs of the interaction strength $g$ yield such different behaviors, the expression \eqref{gse} for the ground state energy of the system implies that the ground state energy does not depend on the sign of $g$. 

\section{Methods \label{methods}}

Given a Hamiltonian describing a quantum system of $N$ indistinguishable bosons, we would like to solve the time independent Schrödinger equation for the ground state energy and wavefunction. To do this, we employ the variational principle; any arbitrary wavefunction $\psi$ must have an expectation value of the energy at least as large as the ground state energy. 
\begin{equation}
    \langle E\rangle_0 \leq \langle E(\theta)\rangle = \frac{\int dX\: \psi^{\dagger}(X, \theta)\hat{H}\psi(X, \theta)}{\int dX \:\psi^{\dagger}(X, \theta)\psi(X, \theta)} \label{varprinciple}
\end{equation}
Here $X$ stands for the positions of the $N$ bosons. Typically, one begins with an ansatz for the wavefunction $\psi(X, \theta)$, where the $\theta$ are its variational parameters and then minimize $\langle E(\theta)\rangle$ with respect to $\theta$ to obtain an estimate for the ground state of the quantum system.\\

\subsection{Neural Network Wavefunction Antsatze \label{nnansatz}}

The quality of the estimate for the ground state energy and wavefunction depends both on the number of variational parameters one uses as well as the particular functional form of $\psi$. Although one could successfully make well-posed guesses as to the correct functional form of the ground state wavefunction based on expectations on the underlying physics of the system, we would like to more systematically search the space of potential ground state wavefunctions. In this context, the application of neural networks is powerful. 

\subsubsection{Neural Networks \label{NNs}}

We define a neural network with $L$ layers as a function which takes $n_1 = N$ inputs and produces $n_L = M$ outputs --- defined in the following way:
\begin{equation}
    \begin{split}
    & I_{i+1} = O_{i}= f(\mathbf{W}_{i} \vec{I_{i}} + \vec{b}_{i}),\:\:1 \leq i \leq L-1 \\
    & O_L = \mathbf{W}_L \vec{I}_L + \vec{b}_L
    \end{split}
\end{equation}
The inputs $I_i$ at layer $i$ are multiplied by an $n_{i+1} \times n_{i}$ matrix of weights $\mathbf{W}_i$ \footnote{For clarity, $n_1 = N$ and $n_L = M$; $n_i$ is the number of nodes at layer $i$.}. A bias vector $\vec{b}_i$ of length $n_{i+1}$ is added before applying an activation function $f$ to each resulting vector element. The output $O_i$ of the $i$-th layer then becomes the input for the $(i+1)$-th layer, where the same procedure is repeated with a different set of weights and biases. While in general, a different activation function can be used at each layer, we choose to keep the same function throughout. To ensure that the neural network produces an output whose range is over all reals, an activation function is \textit{not} used in the final output layer.\\

Neural networks are useful in the context of writing down a variational ansatz for the wavefunction because multilayer neural networks have been shown to be universal function approximators \cite{HORNIK1989359, HORNIK1991251}.
%In other words, a neural network with a modest number of layers and smooth activation functions can approximate any multivariate function and its derivatives up to arbitrary accuracy as the number of nodes in each layer is increased. 
The advantage is that, without needing to change the architecture of the neural network function, we can explore a general space of functions to find the ground state wavefunction by varying the parameters $\theta = \{\mathbf{W}_i, \vec{b}_i\}$. The efficiency of the functional expressivity for many-body wavefunctions, however, is not clear and is a property we seek to explore. We denote our ansatz neural network function as $\mathcal{A}(I_1, ..., I_N)$; it takes in $N$ inputs and returns a single output ($M = 1$). 
\\

\subsubsection{Enforcing Ground State Symmetries \label{symmetries}}

A neural network ansatz for the ground state wavefunction of a system of $N$ indistinguishable bosons should satisfy the following two constraints. Firstly, the wavefunction must obey Bose exchange symmetry.
\begin{equation}
    \psi(..., x_i, ..., x_j, ...) = \psi(..., x_j, ..., x_i, ...)
\end{equation}
where the $x_n$ are the positions of the bosons. Thus we must construct our neural network such that it is fully symmetric under exchange of particle positions. To do this, we look for a bijection between the set of particle positions $\{x_i\}$ and a set of functions of $\{x_i\}$ which are symmetric under $x_i \leftrightarrow x_j$. The simplest such bijection generates the set of elementary symmetric polynomials 
\begin{equation}
    e_i = \sum_{1 \leq j_1 < j_2 < ... < j_i \leq N} x_{j_1}x_{j_2}...x_{j_i}
\end{equation}
For instance, for $N = 3$:
\begin{equation}
    \begin{split}
        & e_1 = x_1 + x_2 + x_3, \\
        & e_2 = x_1x_2 + x_1x_3 + x_2x_3, \\
        & e_3 = x_1x_2x_3.
    \end{split}
\end{equation}
We can see this by noting that the $e_i$ are the coefficients of a degree-$N$ polynomial over an auxiliary variable $t$ whose $N$ roots are the positions $x_i$ --- namely
\begin{equation}
    f(t) = (t-x_1)(t-x_2)...(t-x_N)  
\end{equation}
For ease of implementation however, we note that the Newton-Girard identities define a further bijection between the set of $\{e_i\}$ to the set of power-sums
\begin{equation}
    \xi_i = \sum_{k = 1}^{N}\:x_k^{i}
\end{equation}
Since the set of $\{\xi_i\}$ are also fully symmetric under exchange of $x_i$, we can construct a Bose symmetric neural network function by feeding in the $\{\xi_i\}$ as inputs ($I_i = \xi_i$). In practice, for large $N$, $|\xi_i|$ can easily become extremely large; thus, we normalize the initial particle positions by an approximate width $w$ of the position-space ground state wavefunction \footnote{For example, if the system were a harmonic oscillator with $m = \omega = 1$, we choose $w \approx 2$ such that the majority of the nonzero region of the Gaussian ground state is contained within the selected width.}
\begin{equation}
    \tilde{\xi}_i = \sum_{k=1}^{N} \left(\frac{x_k}{w}\right)^i
\end{equation}
such that the $\tilde{\xi}_i$ are at most $\mathcal{O}(1)$ throughout the search for the ground state wavefunction.
\begin{equation}
    \psi(x_1, ..., x_N) \sim \mathcal{A}(\tilde{\xi}_1, ..., \tilde{\xi}_N)
\end{equation}
The $\{\tilde{\xi}_i\}$ retains information about the original particle coordinates but loses information about their ordering, as desired.\\
%It is important to note that by constructing the neural network wavefunction from the $\{\tilde{\xi}_i\}$, we neither lose information about the original particle coordinates nor fix a particular structure for the function space (we can reconstruct the $\{x_i\}$ given the $\{\tilde{\xi}_i\}$). \\

Secondly, the ground state wavefunction of a system of indistinguishable bosons can be taken to be real and positive. Thus, we constrain our neural network weights and biases to be real-valued. We can also, without loss of generality, write the wavefunction ansatz as
\begin{equation}
    \psi(x_1, ..., x_N) = e^{-\mathcal{A}(\tilde{\xi}_1, ..., \tilde{\xi}_N)}\cdot e^{-\Omega \sum_{i = 1}^{N}\:x_i^2} \label{ansatz}
\end{equation}
Since we are looking for bound ground state solutions of the Hamiltonians models we consider, we want to search the space of wavefunctions which vanish at infinity. We multiply our neural network function by the Gaussian factor to ensure that this property is satisfied even when the neural network function itself may be relatively flat over its inputs for a given set of weights and biases. Because the bosons in the quantum systems we study are confined in a harmonic trap with $m = \omega = 1$, we choose $\Omega = 0.5$. The neural network function $\mathcal{A}$, then, is purely responsible for capturing the complexities of particle interactions. At zero inter-particle interaction strength, $\mathcal{A}$ should minimize trivially to zero everywhere in the ground state.

\subsection{Neural Network Architecture \label{arch}}

To search for the ground state energy and wavefunction of a quantum system, we first construct a neural network wavefunction ansatz of the form in \eqref{ansatz}. For the neural network, we choose to use the continuously-differentiable CELU as our activation function $f$.
\begin{equation}
    \text{CELU}(x) = \begin{cases}
        x & x > 0\\
        e^x - 1 & x \leq 0 
    \end{cases}
\end{equation}
Furthermore, we initialize the real-valued weights and biases of the neural network to be $\mathcal{O}(1)$ and nonzero --- drawn randomly from a normal distribution centered around zero with standard deviation $\sigma = 1$. Using $\langle E\rangle$ as a cost function, we perform gradient descent using the Adam optimizer \cite{Kingma:2014vow} to approach the ground state. We use Adam step sizes of between $10^{-3}$ and $10^{-6}$, decreasing the step size as the minimization approaches closer to convergence. Using relatively large step sizes in certain regions of parameter space close to convergence can lead to being stuck in orbit cycles and exhibit oscillatory behavior in the cost function \cite{nar2018step}. The other hyperparameters of the Adam optimizer describing the decay rates of the moments and ensuring numerical stability are kept at their default values: $\beta_1 = 0.9, \beta_2 = 0.999, \epsilon = 10^{-8}$.\\

Because the integrals involved in obtaining $\langle E\rangle$ and $\langle \partial E /\partial \theta \rangle$ are difficult to compute exactly for a complicated neural network function and for many particles $N$, we estimate the energy and the gradient using Metropolis importance sampling, drawing from the distribution $\pi(X) \sim \psi(X, \theta)^2 / \int dX\: \psi(X, \theta)^2$. However, the Hamiltonians which we study contain potential terms which involve delta functions in position. This creates a zero-overlap problem in the Monte Carlo evaluation of this integral in which none of the sampled position configurations induce a contribution to the energy or gradient from those delta function terms. We introduce a remedy in Appendix \ref{appendb}. Although computing the gradient using Monte Carlo methods is not as precise as an analytic calculation, the stochasticity tends to aid in avoiding being stuck in local minima during the minimization \cite{bottou1991stochastic, AMARI1993185, DBLP:journals/corr/abs-1802-06175}. \\

\begin{figure}[t!]
\begin{minipage}{.48\textwidth}
  %\centering
\includegraphics[width=1.07\textwidth]{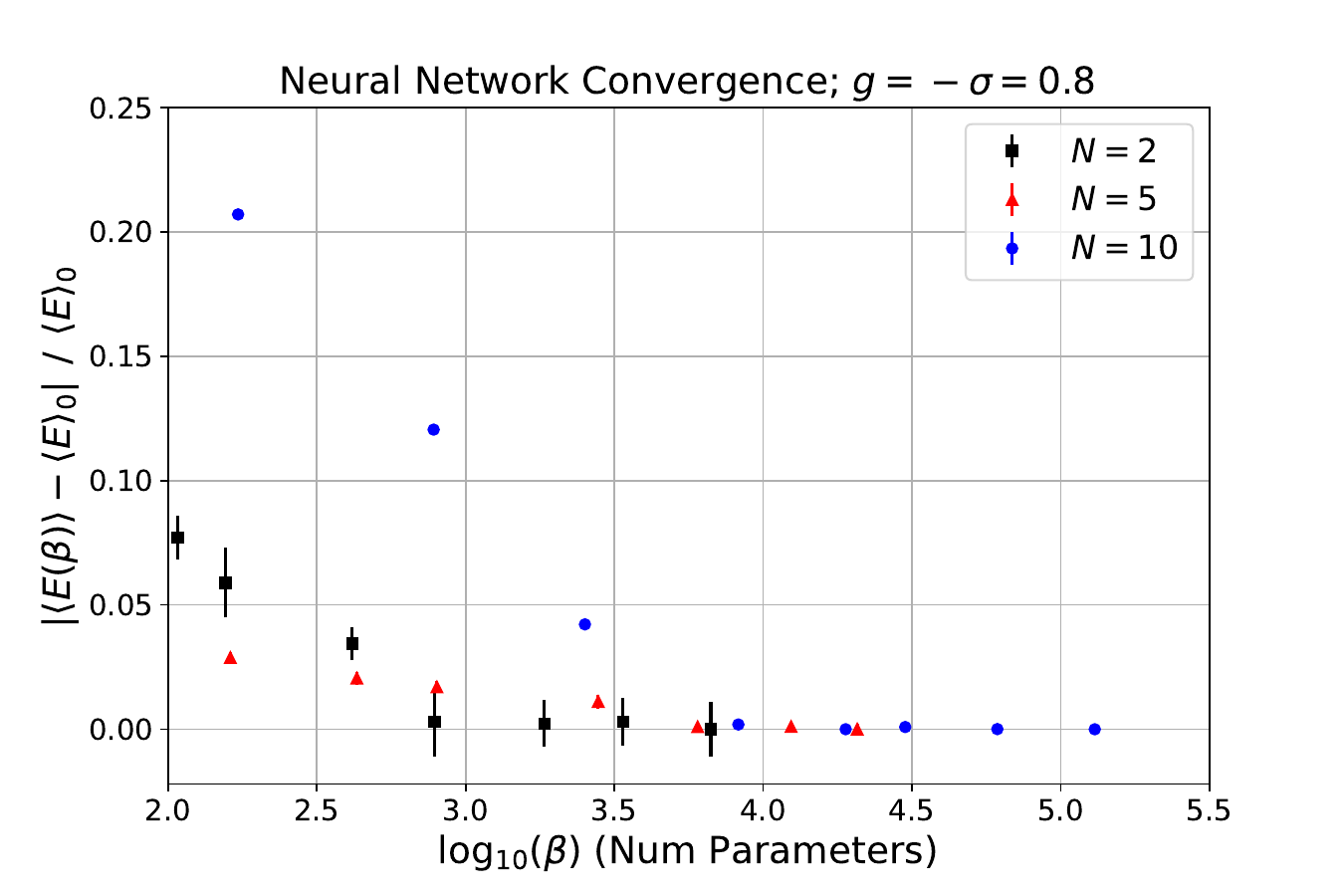}
\caption{\small{Shows the relative error of the ground state energies for the quantum system of $N$ bosons described by the Hamiltonian in \eqref{longham} as the number of neural network parameters $\beta$ is increased. Eight layers were used in the neural network throughout. The statistical error bars from the Monte Carlo calculations on some points are too small to be seen. $\langle E\rangle_0$ here is taken to be the value of the ground state energy computed with the largest $\beta$. \label{fig:conv}}}
\end{minipage}\hspace*{\fill}
\end{figure}

In general, there has not been a systematic solution to choosing an optimal number of layers and nodes of a neural network for any given problem \cite{doi:10.1080/01431160802549278}. Here, we arbitrarily choose to construct our neural network function with between four to ten layers. Although a more careful analysis in choosing the number of layers has yet to be done, we decide to use more layers when solving quantum systems of more particles since we expect that more layers are needed to capture the ground state of a more complicated system. Keeping the same number of layers in solving a particular system, we systematically increase the number of nodes at each layer until we find that our prediction for the ground state energy ceases to decrease within error bars given by the finite Monte Carlo sampling. Fig. \ref{fig:conv} shows convergence in $\langle E(\beta) \rangle$ as we increase the total number of variational parameters (weights and biases) in the neural network. We find that an increasing number of parameters for systems of more particles is needed; however, this dependence appears to be slower than exponential in number of particles.\\

Finally, to perform the neural network computations and gradient descent minimizations, we utilize Google's JAX \cite{jax2018github} package in Python. JAX has built-in automatic differentiation, just-in-time compilation speed-up, and vectorization features which make it an appealing and convenient choice to run neural network computations with. 

\section{Results \label{results}}

To check that our variational neural network wavefunction ansatz \eqref{ansatz} is universal enough to capture the ground state behaviors of many different quantum systems, we use our ansatz to compute the ground states in various phases of the Hamiltonian models described in Sec. \ref{models}. We compare our results to the analytic solutions of the exactly solvable models. We then also explore an unsolved regime of one of these models.\\

\begin{figure}[b!]
\begin{minipage}{.48\textwidth}
  %\centering
\includegraphics[width=1.1\textwidth]{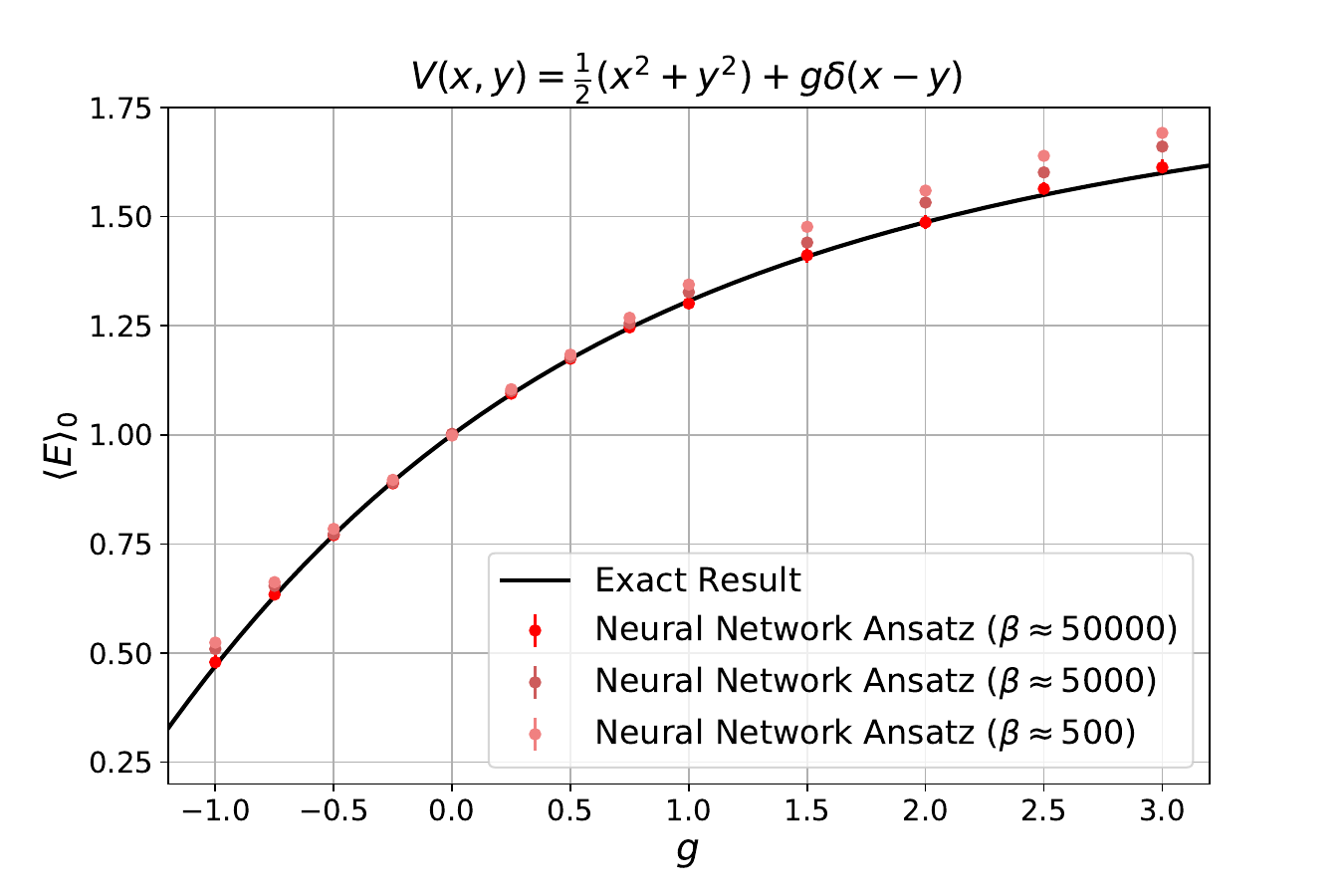}
\caption{\small{Shows the ground state energies of the quantum system described in \ref{coldbosons}. We plot the results computed from our neural network ansatz (red) against the exact results (black) given by Busch et. al. \cite{busch}. The statistical error bars from the Monte Carlo integration of the energy in most cases are too small to be seen. Convergence towards the exact result is seen as the number of parameters $\beta$ in the neural network ansatz is increased. \label{fig:harmLL}}}
\end{minipage}\hspace*{\fill}
\end{figure}

\subsection{Cold Bosons in a Harmonic Trap}

Preliminarily, we first study a one-dimensional system of two indistinguishable bosons ($N = 2$) in a harmonic trap, interacting via a contact potential (\ref{coldbosons}). An exact expression for the ground state energy of the system was given by Busch et. al. \cite{busch}. Using our neural network ansatz, we compute the ground state energies of the system at various interaction strengths $g$. Fig. \ref{fig:harmLL} shows a comparison of the two results. As we increase the number of parameters $\beta$ in the neural network ansatz, we find that, over the wide range of $g$, the neural network ground state energies converge towards agreement with those of the analytic solution to within error bars. The mean values are also consistent with the exact expression to approximately one to two percent precision. Throughout the final calculations for this model, we fixed our neural network to have four layers and roughly $5\cdot 10^{4}$ variational parameters. Although this likely uses more parameters than needed to describe the simpler two-boson system, the results give confidence that our Bose symmetric neural network ansatz, as desired, is able to capture the ground state physics of the different systems without needing to specify a particular functional form for the ansatz or adjusting the architecture of the neural network. 

\begin{figure}[t!]
\centering
\begin{minipage}{.48\textwidth}
\includegraphics[width=1.1\textwidth]{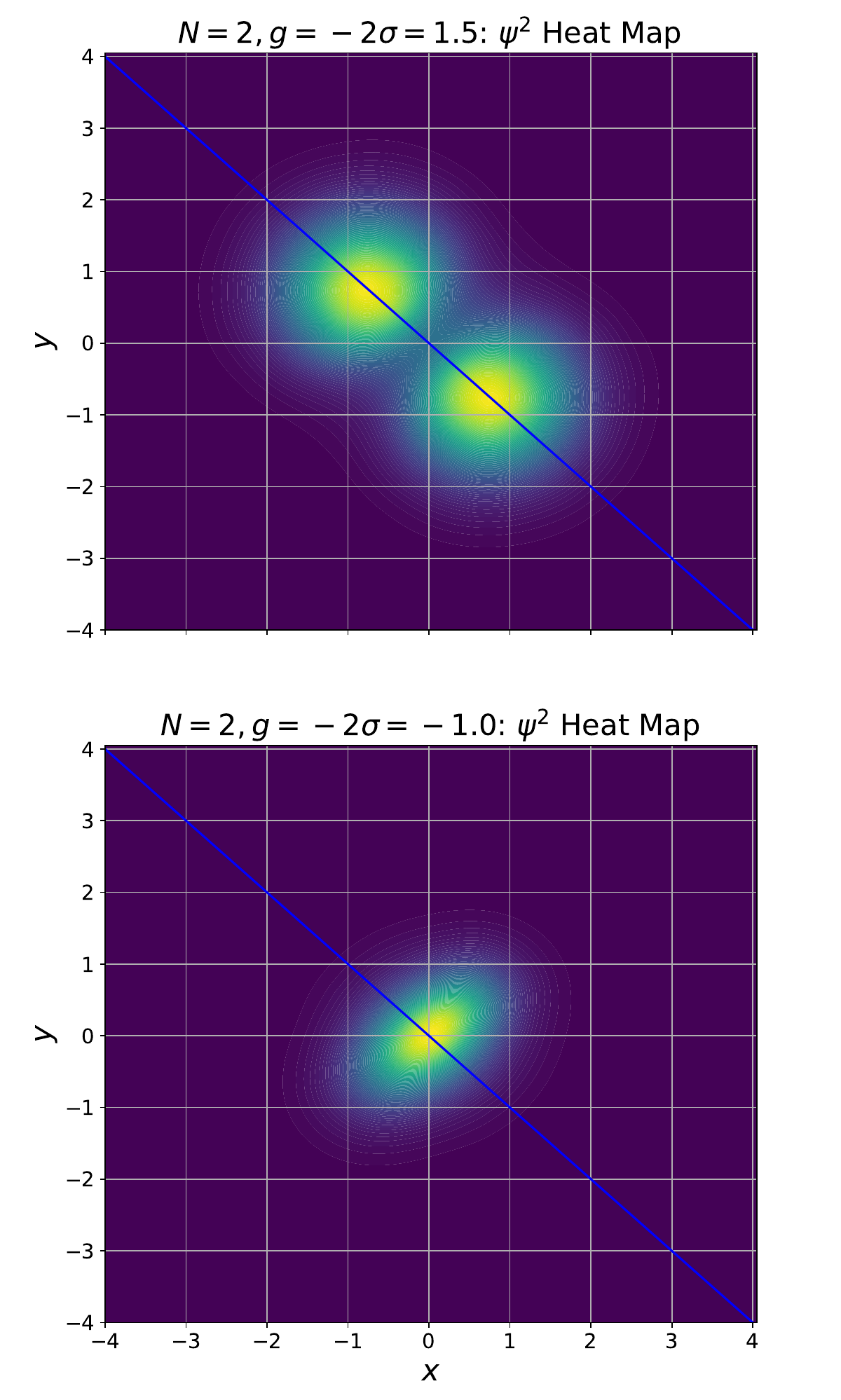}
%\end{minipage}
\caption{\small{Shows ground state behaviors of two bosons interacting with short and long-range interactions described in \ref{longrangeint} in the exactly-solvable regime $\sigma = -g/2$. We plot the position-space probability density; brighter colors indicate a larger value of $\psi(x,y)^2$ and the bosons have a higher probability of being there. The blue lines which run through $y = -x$ are shown to guide the eye. The top shows the system with a repulsive potential ($g = 1.5$) while the bottom shows the system with an attractive potential ($g = -1.0$). \label{fig:N2featuresexact}}}
\end{minipage}\hspace*{\fill}
\end{figure}

\subsection{Trapped Bosons with Short and Long-Range Interactions}

Next, we examine the model in which the bosons interact with a long-range linear potential along with the delta-function potential whose properties are summarized in Sec. \ref{longrangeint}. We are able to use the neural network ansatz to study both the exactly-solvable regime of $\sigma = -g/2$ solved by Beau et. al \cite{PhysRevLett.125.220602} as well as the regime $\sigma = -g$ in which no analytic solution is known.\\

\subsubsection{Exactly-solvable Regime ($\sigma = -g/2$)}

\begin{figure}[t!]
\begin{minipage}{0.48\textwidth}
\includegraphics[width=1.03\textwidth]{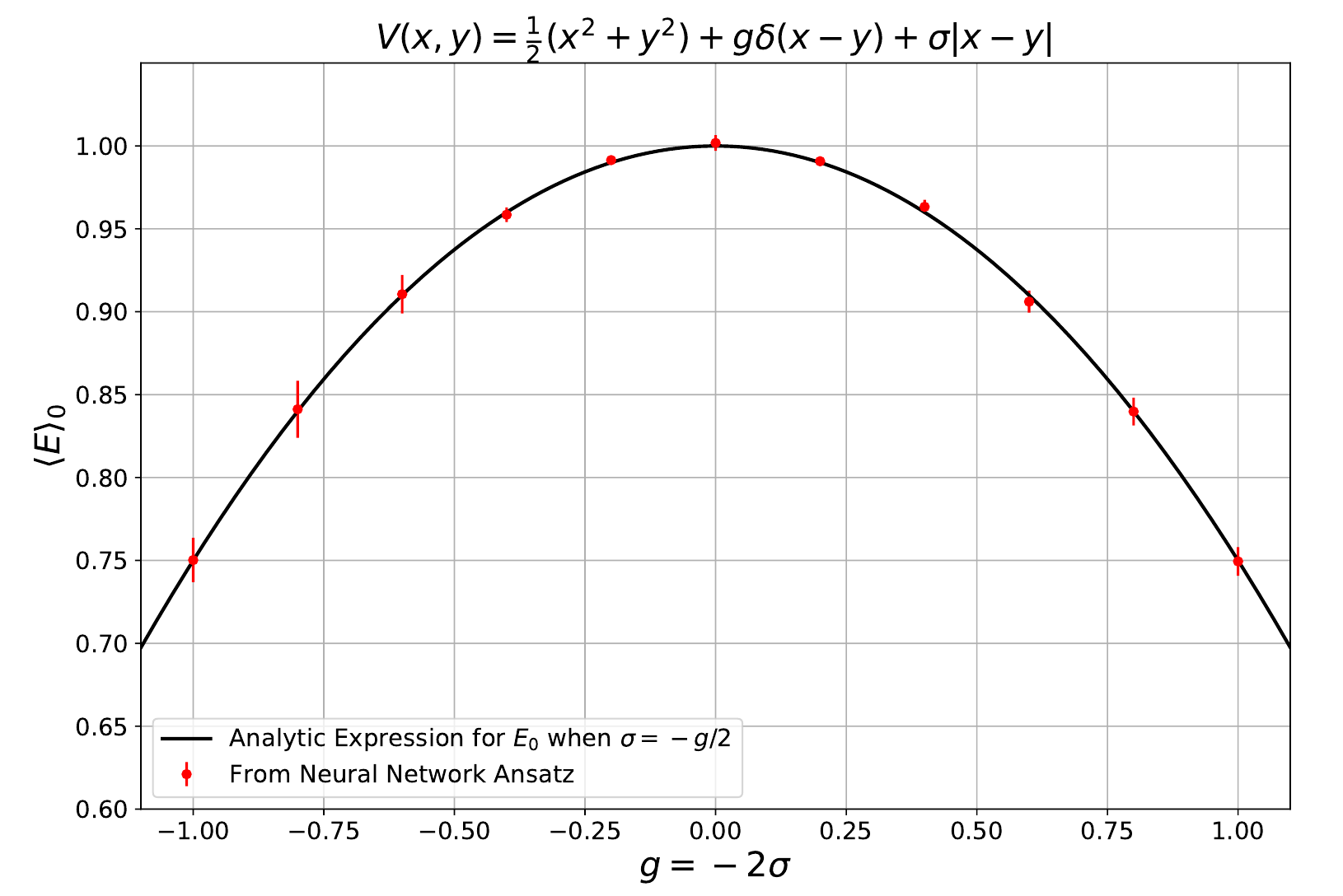}
\centering
%\end{minipage}
\caption{\small{Shows a comparison of the ground state energies of the system \eqref{longham} with two bosons at potential strengths $\sigma = -g/2$. The black line shows the analytic expression for the ground state energy $E_0$ from Beau et. al. \eqref{gse} and the red points are ground state energies computed from the neural network ansatz. \label{fig:N2energiesexact}}}
\end{minipage}\hspace*{\fill}
\end{figure}

We again begin with a simpler system of just two particles. Using our neural network ansatz, we compute the ground state energies and wavefunctions for various values of the coupling in the exactly-solvable regime $g = -\sigma /2$ and compare the results to the analytic solution. Fig. \ref{fig:N2featuresexact} displays the ground state wavefunction of the system while Fig. \ref{fig:N2energiesexact} shows a comparison of the ground state energies. 
We find that the ground states given by the neural network ansatz give the same behavior as described by the exact ground state wavefunction \eqref{gswave}. When $g > 0$ and the interactions are both repulsive, the two bosons separate an equal distance apart from the center of the harmonic trap. Alternatively, when $g < 0$ and the interactions are both attractive, the bosons collapse towards the center of the trap and form a bright soliton. Quantitatively, the ground state energies computed using the variational method also agree with the analytic expression \eqref{gse} to within two-percent statistical errors. The unintuitive quadratic dependence of the ground state energy on $g$ is also confirmed. \\

Having shown the ability of our neural network ansatz to accurately compute ground state wavefunctions and energies of smaller, simpler systems, we further demonstrate the versatility and generalizability of the neural network methodology by computing ground state properties of the quantum system in \eqref{longham} with fifty indistinguishable bosons \footnote{The calculations with $N = 50$, $\beta \approx 2\cdot 10^5$ took roughly 2 min/training step CPU run-time using $8\cdot 10^5$ Monte Carlo steps/training step on a 32-core Lambda workstation.}. We study both a repulsive system ($g = 0.05$) and an attractive system ($g = -0.03$). Using a neural network ansatz with $\beta \approx 2\cdot 10^{5}$, we find that the computed ground state energies fall within one percent of those predicted by the exact $E_0$ \eqref{gse}; the calculations yielded for $g = 0.05$, $E_0^{\text{NN}} = 12.083(72)$ as compared with $E_0 = 11.984375$ and for $g = -0.03$, $E_0^{\text{NN}} = 20.367(76)$ as compared with $E_0 = 20.314375$ .\\
\begin{figure}[t!]
\begin{minipage}{0.48\textwidth}
\includegraphics[width=1.1\textwidth]{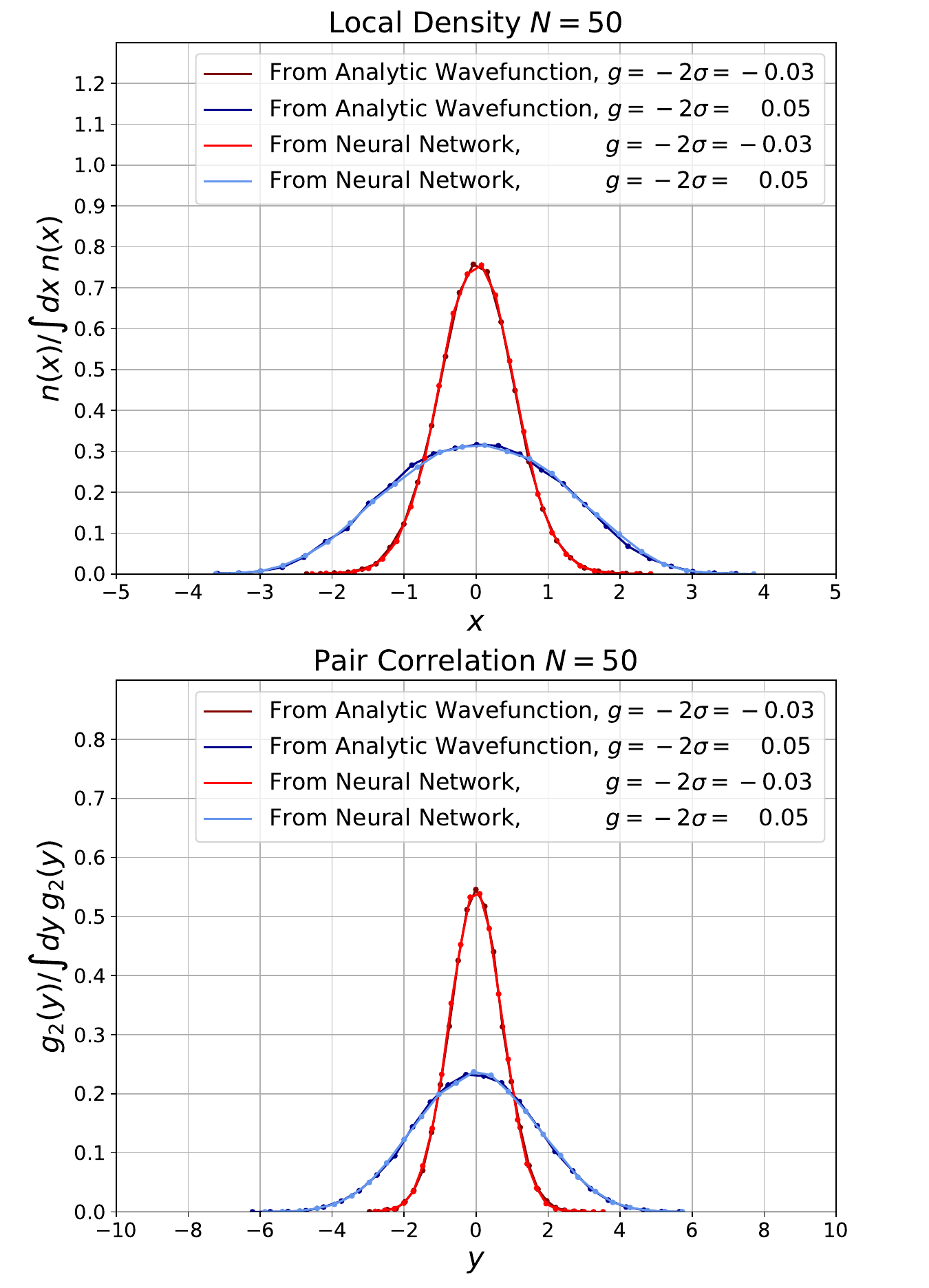}
\centering
%\end{minipage}
\caption{\small{Shows the ground state density functions for the system of fifty bosons with $\sigma = -g/2$. The top panel displays the normalized local density while the bottom displays the normalized pair correlation. The points are heights of the position/distance binning and are linearly interpolated to guide the eye. The dark/light red show the density functions from the analytic/neural network wavefunction respectively in the attractive case while the dark/light blue show the density function sfrom the analytic/neural network wavefunction in the repulsive case. \label{fig:N50densitiesexact}}}
\end{minipage}\hspace*{\fill}
\end{figure}

To comprehensively check of the ground state features of the system, we also calculate the local density profile
\begin{equation}
    n(x) = \int dx_2...dx_N \psi_0(x = x_1, x_2, ..., x_N)^2
\end{equation}
and the density-density correlation function 
\begin{equation}
    g_2(y) = \int dx_1...dx_N\:\delta(y-|x_1-x_2|)\psi_0(x_1, ... ,x_N)^2
\end{equation}
for both the analytic $\psi_0$ \eqref{gswave} and the ground state wavefunction given by the neural network ansatz $\psi_0^{\text{NN}}$ \footnote{The choices $x = x_1$ and $\delta(y - |x_1-x_2|)$ in $n(x)$ and $g_2(y)$ respectively are arbitrary because $\psi_0$ is Bose symmetric.}. The density functions $n(x)$ and $g_2(y)$ are computed by generating histograms of positions $x_1$ and distances $x_1 - x_2$ respectively --- both taken from Monte Carlo samples drawn from the distribution $\pi(X) \sim \psi_0(X)^2/\int dX \psi_0(X)^2$. Fig. \ref{fig:N50densitiesexact} shows the density functions from both $\psi_0$ and $\psi_0^{\text{NN}}$ plotted on top of one another. We find close agreement between the two in both phases of the model ($g > 0$ and $g < 0$). This gives evidence that the neural network ansatz is not only able to provide an accurate estimate for the ground state energy but can also faithfully encode information about the ground state wavefunction. 

\subsubsection{Non-integrable Regime ($\sigma = -g$)}

The variational approach using neural network ansatze is not restricted to computing integrable systems. We also apply the methodology to explore the system of bosons interacting with short and long-range interactions \eqref{longham} with symmetric coupling strengths $\sigma = -g$ for which no exact ground state has been found. We then compare our computed ground state properties to those of the exactly solvable system $\sigma = -g/2$ (see \cite{PhysRevLett.125.220602}). The results show that although the ground state energies behave quantitatively very differently, the general phase structures of the two regimes are similar in nature. \\

\begin{figure}[b!]
\begin{minipage}{0.48\textwidth}
\includegraphics[width=1.1\textwidth]{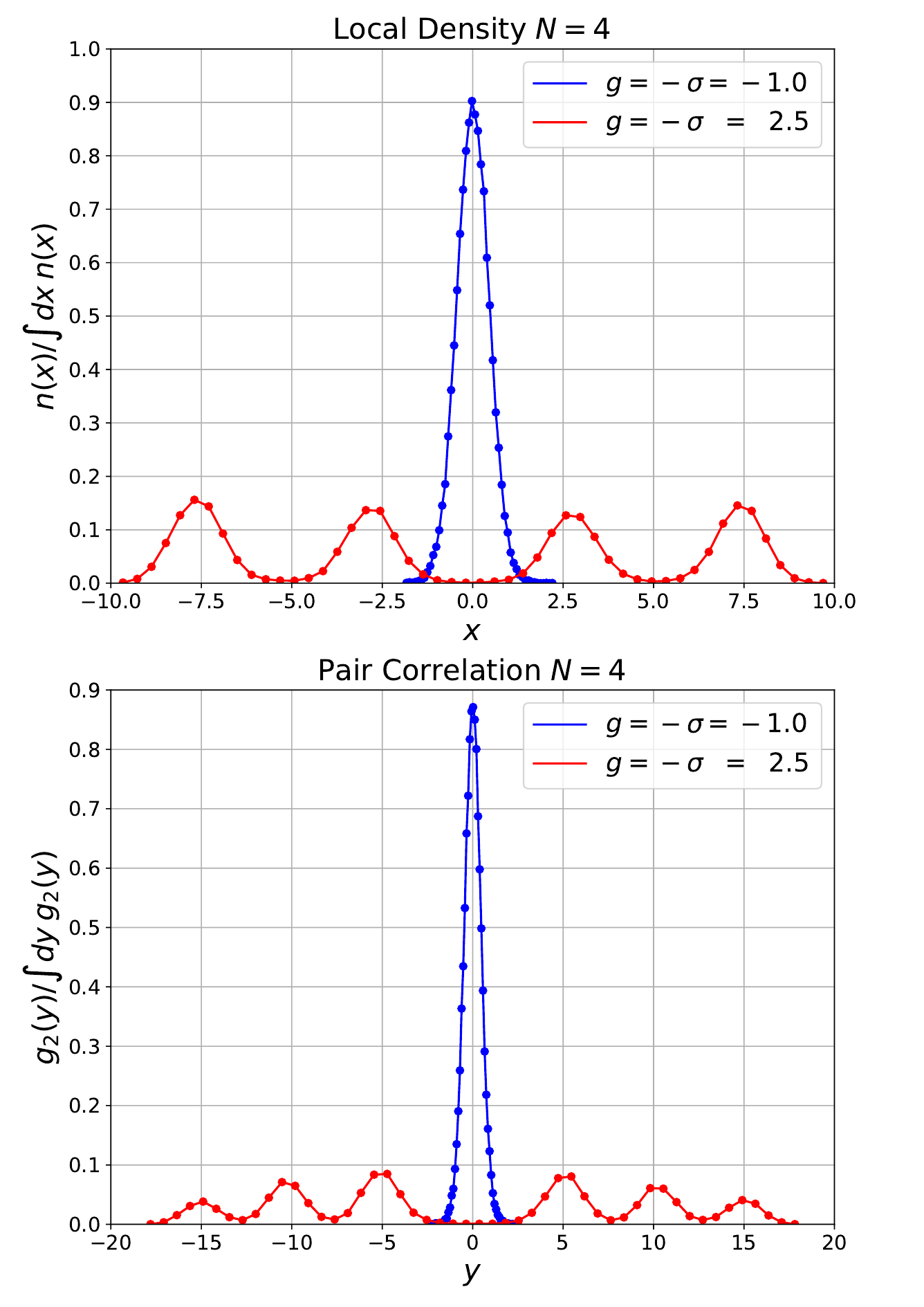}
\centering
%\end{minipage}
\caption{\small{Shows the ground state density functions for the system of four bosons. The top panel displays the normalized local density while the bottom displays the normalized pair correlation. Each panel contains results for the attractive (blue) and the repulsive (red) systems. \label{fig:N4densities}}}
\end{minipage}\hspace*{\fill}
\end{figure}

We begin by looking at a system of four bosons with a strong repulsive coupling $g = -\sigma = 2.5$ and another with an attractive coupling $g = -\sigma = -1.0$. Fig. \ref{fig:N4densities} shows the local density and the pair-correlation function for both four-boson systems. Just as with the exactly-solvable model, a crystal structure is clearly seen in the case of strong repulsion. The local density shows four distinct equally spaced peaks. The pair-correlation function shows that the probability of two particles being in the same position is close to zero. Instead, the correlation function is maximal at distinct separations corresponding to $y = \pm|x_1 - x_2| = nd, n \in \mathds{Z}$ where $d$ is the inter-particle spacing between two nearest neighbors in the crystal. In the case of attraction on the other hand, both the density profile and the pair density peak strongly at $y = 0$; again, all of the particles clump towards the center of the harmonic trap and form a bright soliton \cite{inbook}. \\

\begin{figure}[t!]
\begin{minipage}{0.48\textwidth}
\includegraphics[width=1.1\textwidth]{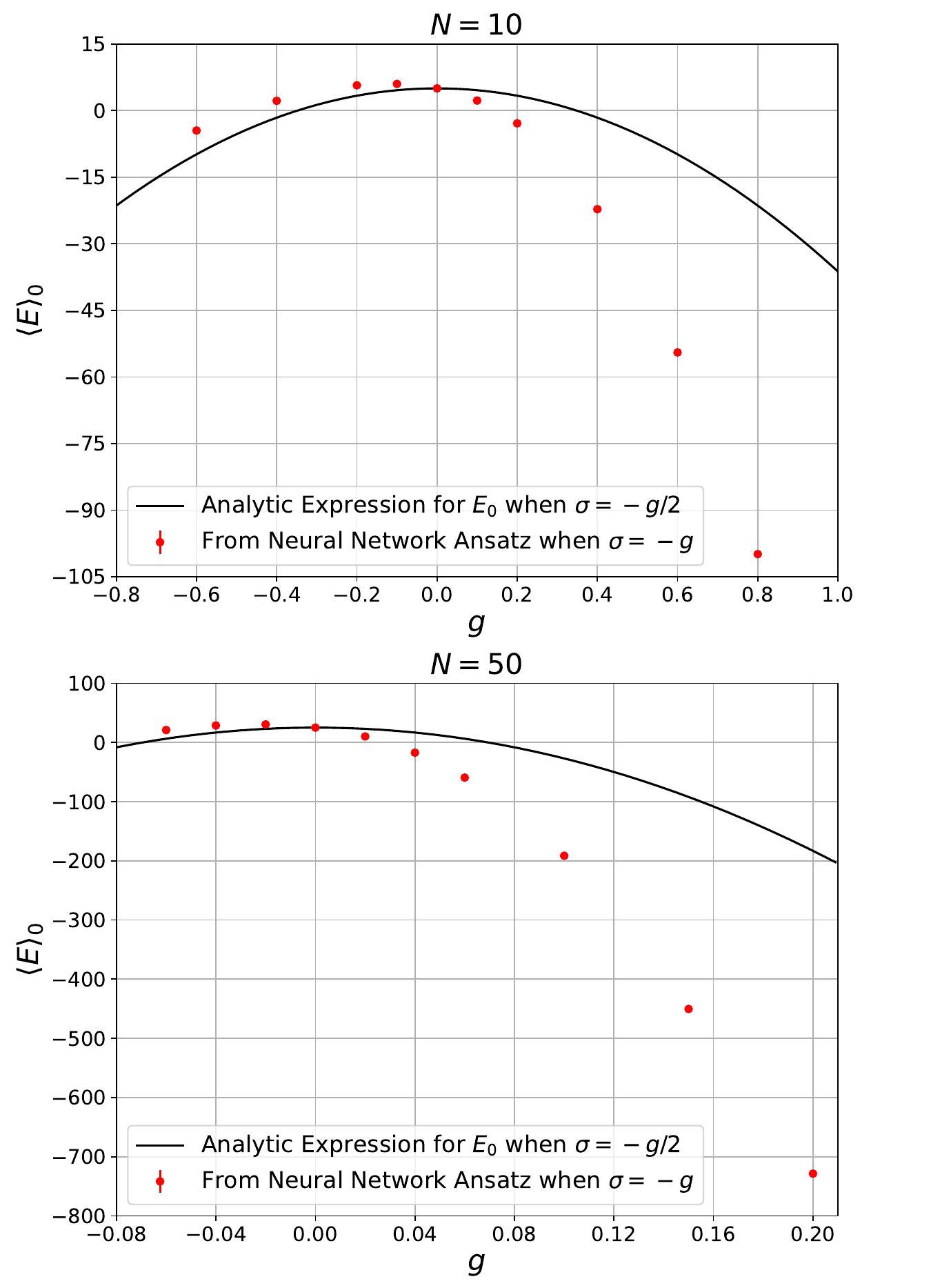}
\centering
%\end{minipage}
\caption{\small{Shows a comparison of the ground state energies of the Hamiltonian \eqref{longham} with N = 10 and 50 bosons. The red points are energies computed from the neural network ansatz at the non-integrable regime $\sigma = -g$. The statistical error are too small to be seen. The black line shows the analytic expression for $E_0$ when $\sigma = -g/2$ for comparison. \label{fig:manyenergies}}}
\end{minipage}\hspace*{\fill}
\end{figure}

\begin{figure}[b!]
\begin{minipage}{0.48\textwidth}
\includegraphics[width=1.1\textwidth]{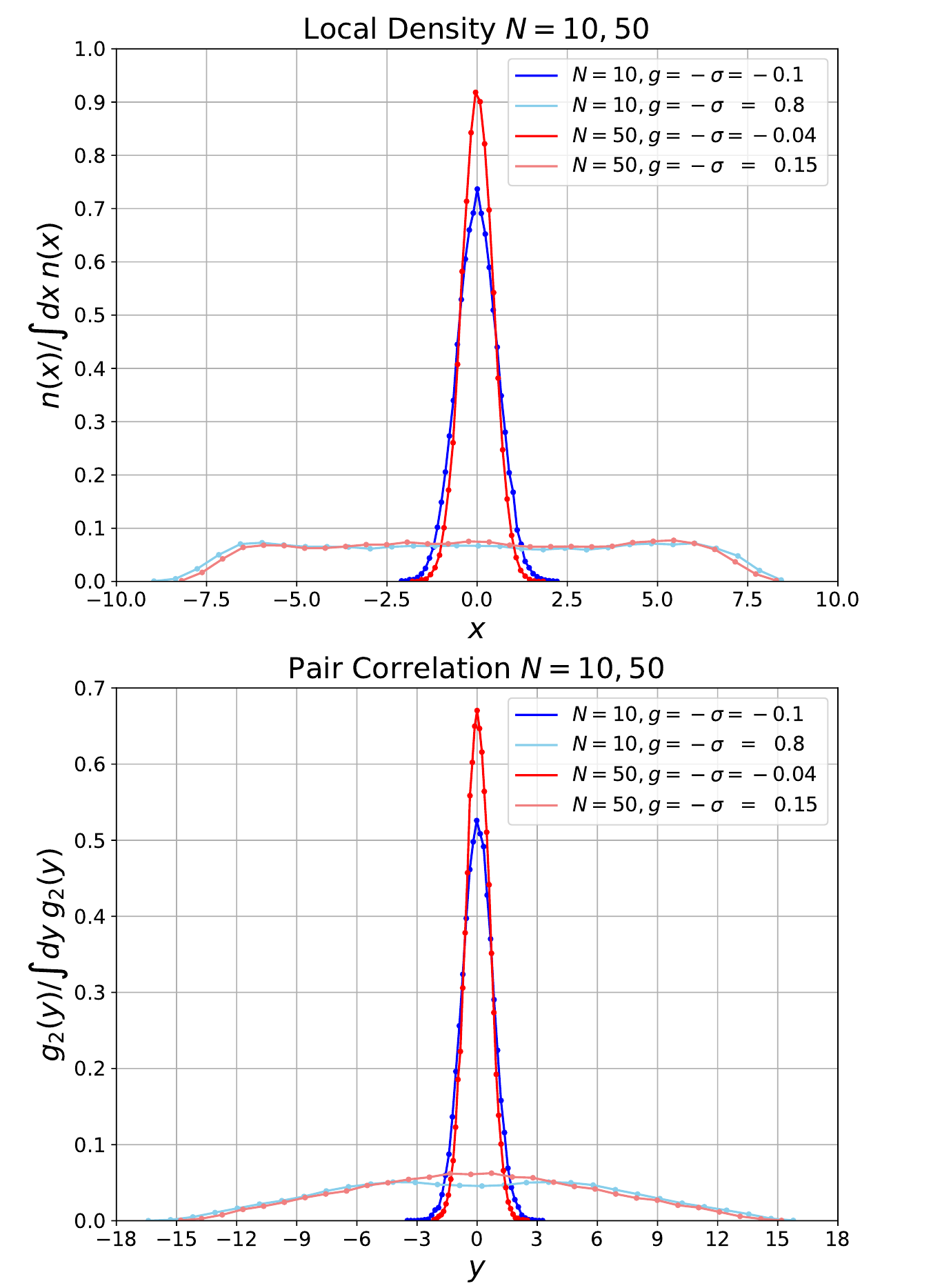}
\centering
%\end{minipage}
\caption{\small{Shows the ground state density functions for systems of ten and fifty bosons. In each panel, the (light) blue shows the (repulsive) attractive system for $N = 10$ and the (light) red shows the (repulsive) attractive system for $N = 50$. \label{fig:manydensities}}}
\end{minipage}\hspace*{\fill}
\end{figure}

We further use our neural network ansatz to search for the ground states of systems of ten and fifty particles at various coupling strengths $g = -\sigma$. Fig. \ref{fig:manyenergies} shows the computed ground state energies as a function of $g$ while Fig. \ref{fig:manydensities} shows the single-particle and pair density profiles. We find that here, the energies exhibit the same asymptotic behavior of $E_0 \rightarrow -\infty$ as $|g| \rightarrow \infty$ as in the exactly-solvable model. However, the ground state energy is no longer independent of the sign of $g$ and changes asymmetrically as the interactions in the system flip from repulsion to attraction. When $g > 0$, the energy drops more rapidly than in the exactly-solvable case while when $g < 0$, the energy of the resulting soliton initially rises before dropping again at higher attractive coupling. \\

The qualitative features, however, again match the behaviors of the system at $\sigma = -g/2$. From the local density profile, we see that as we turn on a repulsive potential ($g > 0$), the bosons begin to spread symmetrically about the center of the harmonic trap. As the repulsion becomes stronger, the particle density shows a mesa-like feature of finite width $w$. This resembles, for example, the behavior of nuclei at saturation density.
%is indicative of a Laughlin-like incompressible liquid \cite{PhysRevLett.50.1395, lieb2018rigidity}
The addition of particles to the system does not change the density but instead increases the droplet size. The two-body correlation for many particles at weak repulsion increases monotonically from the maximal separation $y = \pm w$ inwards towards $y = 0$ since more pairs of particles have smaller rather than larger separations. In the vicinity of zero separation, the probability of finding two particles in the same location is decreased and we expect a slight dip in the pair-correlation function. This is clearly seen in the system with ten particles where the repulsion was stronger; the dip is not seen in the fifty particle system because the weaker repulsion allows the bosons to retain a tendency of remaining close to one another. On the other hand, we again find that if we instead turn on an attractive interaction between the particles, the local density as well as the pair-correlation peak sharply at the center of the harmonic trap and zero separation respectively. This structure is again characteristic to that of a bright soliton.

\section{Conclusions \label{conclusions}}

Neural networks and machine learning techniques provide powerful tools for the Variational Monte Carlo approach to solving many-body quantum systems. We used a neural network to parametrize a variational wavefunction ansatz for the ground state of systems of indistinguishable bosons in one dimension. Then using the ansatz, we were able to successfully numerically compute the ground state energies and wavefunctions of several quantum systems of various qualitative phases, ranging from two to fifty particles. The advantages of using a neural network ansatz include being able to search the most general space of possible ground state wavefunctions without needing to limit to specific functional forms. Neural networks also have convenient scalability; it is straightforward both to increase the number of variational parameters in the ansatz as well as increase the number of particles in the system. The application of such ansatze also does not need to be restricted to solvable systems and can be used to explore properties of non-integrable systems. \\

Many important many-body systems, including condensates, atoms, and nuclei involve indistinguishable particles. Wavefunctions of such systems must satisfy Fermi or Bose exchange symmetry. We advocated and demonstrated that using symmetric functions of the original particle coordinates as inputs to the neural network aptly enforces Bose exchange symmetry in one dimension. The symmetric inputs are chosen to be bijective to the original set of coordinates and so we do not limit the generality of the wavefunction space spanned by the neural network. In addition, we propose a method to deal with delta-function interactions without the need to regularize them.

The use of symmetric inputs to enforce exchange symmetries of the system can be generalized to use to study systems in a higher number of spatial dimensions in a straightforward way. To enforce Bose symmetry in three dimensions, one needs to find a bijection between the set of vector coordinates $\{\vec{r}_i\}, i = 1, ..., N$ and symmetric functions $\{s_j(r_1^x, r_1^y, r_1^z, ..., r_N^x, r_N^y, r_N^z)\}, j = 1, ..., M$ where $M$ may be greater than $3N$. The $\{s_j\}$ must be symmetric in exchanges $\vec{r}_n \leftrightarrow \vec{r}_m$ but not in exchanges of individual coordinate directions (i.e $r_n^x \leftrightarrow r_m^x$). Given the set of $\{s_j\}$, one should be able to reconstruct the set of $\{r_i^x, r_i^y, r_i^z\}$ including the fact that particular sets of three coordinates specify the position of individual particles. To write a Fermi anti-symmetric function in three dimensions, one theoretically simply needs to take a single Slater determinant of functions $\phi_i(\vec{r}_j;\{\vec{r}_{j/}\})$, where $\phi_i$ is \textit{Bose} symmetric under exchange of any $\vec{r}_n \rightarrow \vec{r}_m, m, n \neq j$ \cite{Pfau_2020}. Thus if one finds a bijection to a set of symmetric functions necessary for enforcing Bose symmetry in three dimensions, one can use the same bijection to enforce Fermi symmetry. Work to find such a mapping is being pursued.

\begin{acknowledgments}

We thank Andrei Alexandru and Scott Lawrence for helpful discussions and comments. This work was supported in part by the U.S. Department of Energy, Office of Nuclear Physics under Award Numbers DE- SC0021143, DE-FG02-93ER40762.

\end{acknowledgments}

\appendix

\section{Importance Sampling for Delta-function Observables \label{appendb}}

Consider the Lieb-Liniger Hamiltonian \cite{PhysRev.130.1605, PhysRev.130.1616} for a system of $N$ indistinguishable bosons
\begin{equation}
    \hat{H} = \sum_{i = 1}^{N} -\frac{1}{2}\frac{\partial^2}{\partial x_i^2} + \sum_{i < j}^{N}\:g\delta(x_i - x_j)
\end{equation}
and a positive and real wavefunction 
\begin{equation}
    \psi(X) = e^{-A(x_1, ..., x_N)}
\end{equation}
We would like to compute the energy associated with the wavefunction.
\begin{equation}
    \langle E\rangle = \frac{\int dx_1..dx_N\:\psi \hat{H}\psi}{\int dx_1...dx_N\:\psi^2} = K + V
\end{equation}
where
\begin{equation}
\begin{split}
    &K = \frac{\int dx_1...dx_N\:e^{-2A(X)}\sum_i\:\frac{1}{2}\left(\frac{\partial^2A}{\partial x_i^2} - \left(\frac{\partial A}{\partial x_i}\right)^2\right)}{\int dx_1...dx_N\:e^{-2A(X)}} \\\\
    &V = \frac{\int dx_1...dx_N\:e^{-2A(X)}\sum_{i<j}g\delta(x_i-x_j)}{\int dx_1...dx_N\:e^{-2A(X)}}
\end{split}
\end{equation}
The kinetic term $K$ can be computed easily using importance sampling
\begin{equation}
    K = \Biggl\langle\: \sum_{i=1}^{N}\:\frac{1}{2}\left(\frac{\partial^2A}{\partial x_i^2} - \left(\frac{\partial A}{\partial x_i}\right)^2\right)\:\Biggr\rangle_{\psi^2}
\end{equation}
where $\langle ...\rangle_{\psi^2}$ denotes an average over samples drawn from the distribution
\begin{equation}
    \Pi_{\psi^2}(x_1,...x_N) = \frac{e^{-2A(X)}}{\int dx_1...dx_N\:e^{-2A(X)}}
\end{equation}
Using the same importance sampling technique to compute the potential term $V$ leads to a zero-overlap problem as described in Sec. \ref{arch}. Instead we restructure the integral involved in $V$ to become amenable to importance sampling using the following manipulations.\\

We first explicitly integrate out the $x_2$ coordinate using the delta functions. The choice of $x_2$ is arbitrary due to the Bose exchange symmetry of the wavefunction. 
\begin{equation}
\begin{split}
    &V = g\frac{N(N-1)}{2}I\\
    &I = \frac{\int dx_1dx_3...dx_N\:e^{-2A(\tilde{X})}}{\int dx_1dx_2...dx_N\:e^{-2A(X)}}
\end{split}
\end{equation}
where $\tilde{X}$ denotes the set of positions where $x_2$ is replaced by $x_1$ (i.e $\{x_1, x_1, x_3, ..., x_N\}$). We use the Bose symmetry of the wavefunction to eliminate the sum over pairs of particles. Next, we rewrite the integral $I$ as
\begin{equation}
    I = \frac{\int dx_1...dx_N \frac{\exp(-2A(\tilde{X}))}{\exp(-2A(X))}\mathcal{D}(x_2)\exp(-2A(X))}{\int dx_1...dx_N\:\exp(-2A(X))}
\end{equation}
where $\mathcal{D}(x_2)$ is any function such that 
\begin{equation}
    \int_{-\infty}^{\infty}dx_2\:\mathcal{D}(x_2) = 1
\end{equation}
In practice, we choose a Gaussian form
\begin{equation}
    \mathcal{D}(x_2) = \frac{1}{\alpha\sqrt{\pi}}e^{-x_2^2/\alpha^2}
\end{equation} \\
The parameter $\alpha$, which is different for each different $\psi$, is chosen arbitrarily such that $\mathcal{D}(y_{*}) \approx 10^{-5}$ where $y_{*}$ is the maximum value of $|x_2|$ over many position samples taken from $\Pi_{\psi^2}(X)$. In this way, $\mathcal{D}(x_2)$ is neither too wide such that we emphasize unimportant regions of $\psi^2$ not sampled by the Monte Carlo method nor too narrow such that we don't capture all of the important regions of $\psi^2$.\\

Now, we can also compute $V$ by drawing samples from $\Pi_{\psi^2}(X)$, just as we did for $K$.

\begin{equation}
    V = g\frac{N(N-1)}{2}\Bigl\langle \: \frac{e^{-2A(x_1, x_1, x_3, ..., x_N)}}{e^{-2A(x_1, x_2, ..., x_N)}}\mathcal{D}(x_2)\Bigr\rangle_{\psi^2}
\end{equation}\\

We note that this method of avoiding the zero-overlap problem occasionally produces large error estimates and fails to converge when $e^{-2A(X)}$ is small for many different sampled $X$. However, the method is still reliable because this is precisely the region of $e^{-2A(X)}$ that is sampled the least by the Metropolis algorithm. 

%\newpage
\bibliographystyle{apsrev4-1}
\bibliography{nn.bib}

\end{document}